\documentstyle[aps,prb,twocolumn]{revtex}
\begin{document}
\title{Reentrant spin glass behavior in a layered manganite La$_{1.2}$Sr$_{1.8}$Mn$%
_2$O$_7$ single crystals}
\author{S. H. Chun\cite{byline}, Y. Lyanda-Geller, and M. B. Salamon}
\address{Department of Physics and Materials Research Laboratory, University of\\
Illinois at Urbana-Champaign, Urbana, Illinois 61801-3080}
\author{R. Suryanarayanan, G. Dhalenne, and A. Revcolevschi}
\address{Laboratoire de Physico-Chimie de l'Etat Solide, CNRS, UMR 8648, Bat.414\\
Universit\'{e} Paris-Sud, 91405 Orsay, France}
\date{\today}
\maketitle

\begin{abstract}
We report here a detailed study of AC/DC magnetization and
longitudinal/transverse transport properties of La$_{1.2}$Sr$_{1.8}$Mn$_{2}$O%
$_{7}$ single crystals below $T_{c}$ = 121 K. We find that the resistivity
upturn below 40 K is related to the reentrant spin glass phase at the same
temperature, accompanied by additional anomalous Hall effects. The carrier
concentration from the ordinary Hall effects remains constant during the
transition and is close to the nominal doping level (0.4 holes/Mn). The spin
glass behavior comes from the competition between ferromagnetic double
exchange and antiferromagnetic superexchange interactions, which leads to
phase separation, i.e. a mixture of ferromagnetic and antiferromagnetic
clusters, representing the canted antiferromagnetic state.
\end{abstract}


\section{Introduction}

Recent investigations of nearly-cubic doped perovskite manganites that show
colossal magnetoresistance (CMR) reveal the importance of a delicate balance
among charge, spin, lattice, and orbital degrees of freedom~\cite{review}.
It is also clear that effective dimensionality of a system plays an
important role, as the CMR effect becomes larger in layered manganites,
particularly the $n$=2 member of the Ruddlesden-Popper series (La,Sr)$_{n+1}$%
Mn$_{n}$O$_{3n+1}$, compared to the $n$=$\infty $ case, La$_{1-x}$Sr$_{x}$MnO%
$_{3}$. The ferromagnetic transition temperature $T_{c}$ is, however,
considerably lower ($\sim $120 K) in the former.\cite{Moritomo} Because of
the reduced dimensionality, the balance between ferromagnetic (FM) double
exchange and antiferromagnetic (AFM) superexchange interaction between Mn
ions is more subtle. Therefore, slight changes in doping lead to
significantly different magnetic ground states. For example, Kubota et al.
reported that the ferromagnetic easy axis changes from $c$ axis ($x$ 
%TCIMACRO{\TEXTsymbol{<} }%
%BeginExpansion
\mbox{$<$}%
%EndExpansion
0.32) to $ab$ plane ($x$ 
%TCIMACRO{\TEXTsymbol{>} }%
%BeginExpansion
\mbox{$>$}%
%EndExpansion
0.32) and that additional doping results in a canted antiferromagnetic
state, with the canting angle increasing with doping, saturating at 180$%
^{\circ }$ at $x$ = 0.48.\cite{Kubota} For La$_{1.2}$Sr$_{1.8}$Mn$_{2}$O$%
_{7} $ ($x$ = 0.4), neutron scattering studies show that, in the ground
state, ferromagnetic and antiferromagnetic features coexist~\cite{Hirota}
which can be interpreted as either a canted AFM state or phase separated
clusters.\cite{Moreo} We may expect, then, some related effects in bulk
transport and/or magnetic properties.

Here, we report a detailed study of the magnetic and transport properties of
La$_{1.2}$Sr$_{1.8}$Mn$_{2}$O$_{7}$ single crystals below $T_{c}$ = 121 K.
Our data show that the resistivity upturn below 40 K, typically seen in
previous studies, is related to a transition in the magnetic subsystem from
ferromagnetic to a ``disordered canted'' state. The possibility of a
transition from the ferromagnetic state to one of several canted phases was
discussed theoretically in the early paper by de Gennes.\cite{Gennes} More
recently, the phase diagram of manganites in the presence of the competition
between ferromagnetic and antiferromagnetic interactions has received
renewed attention.\cite{Arovas,Golosov} This competition can lead to phase
separation into a system with FM clusters or to a canted antiferromagnetic
state. Our study of AC/DC magnetization indicates that the magnetic ground
state of the layered system, in the temperature range where the resistivity
upturn occurs, shows characteristics of a reentrant spin glass phase.
Furthermore, we observe that the transport properties in this regime are
characterized by additional anomalous Hall effects.

\section{Experiment}

Single crystal rods of La$_{1.2}$Sr$_{1.8}$Mn$_{2}$O$_{7}$ were grown by the
floating-zone method using a mirror furnace. The composition of the crystals
as revealed by EDX analyses was very close to that of the starting material
used as polycrystalline feed rods in the growth furnace~\cite{sample}. For
transport measurements a thin bar shaped sample was cut from one of the rods
such that the surface is perpendicular to the $c$-axis. Contact pads were
made by Au sputtering and Au wires were attached using silver paint. We
adopted a low frequency ac method for the measurements. AC/DC magnetizations
of the same sample were measured by SQUID magnetometers equipped with an ac
susceptibility setup.

The in-plane resistivity $\rho _{xx}$ and low field ($H$ = 5 Oe // $ab$%
-plane) dc magnetization $m$ of our crystal are shown as a function of
temperature in Fig.~1. As typically seen in other studies, the resistivity
drops sharply by more than two orders of magnitude at the ferromagnetic
transition temperature ($T_{c}=$ 121 K). Application of a magnetic field
drastically reduces $\rho _{xx}$ over a wide range of temperatures both
below and above $T_{c}$, resulting in an MR ratio ($[\rho ($0 T$)-\rho ($7 T$%
)]/\rho ($7 T$)$) as large as 14000 \% at $T_{c}$. An interesting feature is
presence of a resistivity minimum at $T_{min}=$ 40 K (without magnetic
field). Okuda et al. ascribed this minimum to the existence of impurities or
disorder in similar materials (x = 0.35).\cite{Okuda} However, when magnetic
field is applied, the resistivity minimum moves to higher temperatures
compared to the resistivity minimum at zero magnetic field. Furthermore,
there is a significant difference between the zero-field-cooled (ZFC) and
field-cooled (FC) magnetizations (Fig. 1)~\cite{intergrowth,Potter}. The
magnetization studies imply that the ground state is not a simple
ferromagnet, but in fact may indicate a possible spin-glass transition.

In order to clarify the origin of the ZFC-FC hysteresis, we measured the ac
magnetic susceptibility of our sample. The remanent DC magnetic field from
the superconducting magnet was nulled below 1 mOe before the measurement.
Figure 2(a) shows the temperature and the frequency dependence of the ac
susceptibility ($H_{ac}=$ 1 Oe). What we expect from a ferromagnet is
similar to the dashed line in the figure, which is the ac susceptibility of
\ a ferromagnetic La$_{0.7}$Ca$_{0.3}$MnO$_{3}$ single crystal ($T_{c}$ was
scaled for comparison). Compared to this, our crystal shows a totally
different behavior. After the ferromagnetic transition on cooling, the real
part of the ac susceptibility ($\chi _{ac}^{\prime }$) decreases again below
50 K, a typical shape for a reentrant spin glass transition.\cite{Coles} The
appearance of frequency dependence only below 50 K strongly supports the
glassy nature of the magnetic state in this region.

The spin-glass phase is characterized by sluggish response to changes in
external field and by slow relaxation of remanent magnetic moments. A useful
method to deduce the glass transition temperature and the relaxation rate is
to measure the temperature and time dependence of the thermoremanent
magnetization (TRM).\cite{Guy} The TRM is the dc magnetic moment measured
after field-cooling the sample from the paramagnetic state and removing the
field at a certain temperature. The inset of Fig. 2(b) shows the time
dependence of the TRM measured up to 6 hours at 5 K after cooling the
crystal under $H$= 100 Oe from 300 K. The apparently logarithmic time
dependence is typical of spin glasses. The normalized relaxation rate, which
is the slope of relaxation divided by the initial magnetic moment $dM/d\ln
t/M_{init}$, was 0.015 at 5 K. This value is much smaller than that found,
for example, in the {\it Au}Fe system~\cite{Guy}, which implies smaller
barrier height. However, the detailed temperature dependence of the
normalized relaxation rate is required for an accurate comparison.

The temperature dependence of TRM, shown in the main panel of Fig. 2(b), is
a good measure of the glass transition temperature $T_{f}$. The data were
taken on heating the sample right after removing $H$ at 10 K from the
field-cooled state. As temperature increases, the TRM drops rapidly and
disappears at 40 K, which coincides with the minimum temperature of $\rho
_{xx}$, $T_{min}$ (the small negative moments above $T_{min}$ are due to the
remanent field of the magnet itself about 3 Oe). This confirms that the
longitudinal resistivity upturn is indeed coincident with a change in the
magnetic state. The correspondence between $T_{f}$ and $T_{min}$ is rather
unusual, because in metallic spin glasses the effect of magnetic moment
freezing is barely seen in resistivity measurements. It is the
double-exchange interaction in this system which induces the strong
correlation between magnetism and electronic transport. Application of a
high magnetic field reduces $T_{f}$ significantly as in conventional spin
glasses, and at the same time, makes the system more conductive. The MR is
still large and even increases again below $T_{min}$, whereas there is
nearly no MR in 3D perovskite manganites at low temperatures.

\section{Magnetic state and transport properties of the system.}

De Gennes in his classic paper~\cite{Gennes} proposed that the magnetic
subsystem in the presence of double exchange interaction can undergo two
transitions. One of these transitions is the usual ferromagnet-to-paramagnet
transition at high temperatures while the other, from ferromagnetic to a
canted ferromagnetic, a helical or a disordered spin state, \cite
{transitions} occurs at lower temperatures due to competition between double
exchange and direct antiferromagnetic interactions. Canted-like spin
arrangements minimize the sum of double-exchange energy of spins, $%
E_{d}\propto \cos {\theta /2}$, where $\theta $ is the angular deviation
between spins, and (antiferromagnetic) superexchange energy, $E_{ex}\propto
\cos {\theta }$. At finite temperatures orientation fluctuations of
individual spins cause the mean interaction energies $\langle E_{d}\rangle
\propto \langle \cos {\theta /2}\rangle $ and $\langle E_{ex}\rangle \propto
\langle \cos {\theta }\rangle $ to decrease at different rates. Thus, the
two components in the effective local field change differently with
temperature, leading to decrease of the canting angle, and eventually to
ferromagnetic ordering. However, magnetic fields tend to align spins and to
decrease fluctuations of their orientation, therefore leading to an increase
in the temperature of the canted-state-to-ferromagnet transition.

We now turn to the results of our resistivity measurements (Fig.1) that show
the upturn in the temperature dependence of the resistivity occurring at 40
K in the absence of magnetic field. When magnetic field is applied, the
upturn in the temperature dependence of the resistivity occurs at higher
temperatures (Fig.1). We suggest that the temperature of the upturn in
resistivity is the temperature at which the system begins to cross over to a
de Gennes canted-like state. In crystals with high doping level the
ferromagnet-to-canted-state transition is feasible when the hopping
amplitude which determines the double-exchange interaction is sufficiently
small. In layered systems, the amplitude of hopping between layers is indeed
much smaller than hopping amplitude in cubic manganites, as demonstrated by
strong anisotropy of resistivity in the whole temperature range. We believe
that the resistivity minimum in La$_{1.2}$Sr$_{1.8}$Mn$_{2}$O$_{7}$ is {\it %
intrinsic} and is related to the change in magnetic ground state. Indeed, a
transition from ferromagnetic order to canted-like states should influence
the resistivity of the system. In the simple two-sublattice canted
arrangement, spins are aligned within planes, but form fixed angle with
spins in neighboring plane. When double exchange interaction regulates the
charge carrier motion, the probability of charge carrier to propagate in the
canted-like arrangement is significantly modified compared to isotropic
ferromagnetic arrangement, and the character of charge carrier motion
changes significantly, both because of self-trapping due to local spin
distortions~\cite{Gennes} and because they become more confined to the
two-dimensional planes. It is important to note that an increase in the
resistivity in this case is essentially independent of its mechanism. On one
hand, for example, a metallic system in the diffusive, low temperature
regime might undergo a reentrant transition to an insulating state due to he
effects of quantum interference and electron-electron interactions in the
presence of disorder~\cite{Altshuler}. On the other hand, in the regime of
hopping conduction between localized states~\cite{Efros}, charge carrier
confinement to the two-dimensional planes significantly suppresses charge
percolation. In this regard, we remark that similar temperature dependence
has been observed for both in-plane and c-axis resistivity.

The increase in resistivity must also manifest itself if the spin-glass
transition is between ferromagnetic ordering and a state exhibiting a phase
separation; i.e., with coexisting ferromagnetic and antiferromagnetic
regions. Such a state, in fact, is most likely to occur, as noticed in both
experimental and theoretical studies.\cite{Arovas,Wollan} In our case of
layered manganites, a simple canted arrangement is less likely to manifest
itself, because the system is intrinsically anisotropic, and charge carrier
confinement to two-dimensional planes exists even in the absence of canted
spin arrangements. However, inhomogeneity resulting from phase separation
results in a pronounced resistivity increase. It is noteworthy that the
magnetic field cannot simply change the relative proportions of the two
regions without a decrease in the phase separation temperature. We also
note, that a system with coexisting ferromagnetic and antiferromagnetic
regions is quite likely to be characterized by a spin-glass like response.

Let us now turn our attention to the Hall effect. Recent studies show that
anomalous Hall effects (AHE) in CMR materials depend strongly on the local
magnetic moment texture, another consequence of double exchange.\cite
{ChunPRL,Yuli} The transition into a canted phase, or the freezing of
magnetic moments should also be manifest in the AHE. Figure 3(a) shows the
field dependence of the transverse (Hall) resistivity $\rho _{xy}$ at
several temperatures around $T_{f}$ (the magnetic flux density $B$ was
corrected by the sample demagnetization factor). The overall behavior is the
same for all temperatures below $T_{c}$. As in typical ferromagnetic metals, 
$\rho _{xy}$ can be written as $\rho _{xy}\left( B,T\right) =R_{0}\left(
T\right) B+\mu _{0}R_{S}\left( T\right) M\left( B,T\right) $, where $%
R_{0}\left( T\right) $ and $R_{S}\left( T\right) $ are the coefficients of
the OHE and the AHE, respectively.\cite{Hurd} Once the magnetization becomes
saturated at higher field (about 2 T in this system), $\rho _{xy}$ increases
linearly owing to the positive, hole-like ordinary Hall effect (OHE) $R_{0}$%
. \ In the free electron approximation, the effective charge carrier density
per Mn ($n_{eff}$) can be calculated from $n_{eff}=V/eR_{0}$, where $V$ is
the volume of unit cell containing one Mn atom. As we can see in Fig. 3(b),
up to 75 K, $n_{eff}$ is constant, 0.36$\pm $0.02 holes/Mn, which is very
close to the nominal doping level (0.4 holes/Mn) and, in particular, is
insensitive to the transition at $T_{f}$.\cite{compare3D,ChunBrief}
Therefore, as concluded above, a mobility change causes the resistivity
minimum. Applying a traditional metallic model for charge carrier transport
in our system has to be done with caution. The reason is that estimates
using a Drude-like model indicate that the resistivity of layered manganites
is near or above the maximal Ioffe-Regel-Mott metallic resistivity
throughout the whole range of temperatures, including the region of the
resistivity minimum and the resistivity upturn, for both in-plane and, of
course, c-axis resistivity. However, we observe no temperature dependence
which would indicate hopping conductivity of charge carriers between
localized states. Most probably, the conductivity can be qualitatively
rendered as metallic, with low magnitude due to magnetic and charge
inhomogeneities and strong scattering by magnetic domain boundaries. This
point of view finds additional support from the temperature dependence of
conductivity in the low-temperature regime (Fig. 4). Similarly to~the work
of Okuda et al.,\cite{Okuda} we observe a square-root temperature dependence
of the conductivity. Such a behavior in 3D metallic system at low
temperatures (although significantly smaller in magnitude than we observe)
would indicate electron-electron corrections to conductivity due to
disorder~ \cite{Altshuler}. Similar behavior may occur in layered systems. 
\cite{Abrikosov} However, the relative magnitude of the observed
magnetoconductivity changes and low magnitude of the absolute value of the
conductivity preclude any quantitative analysis.

Turning to $\left| R_{S}\right| ,$ as extracted from the low-field behavior
of $\rho _{xy},$ Fig. 3b shows it to have a minimum close to $T_{min}$. One
may expect this because it is known that $R_{S}$ correlates with $\rho _{xx}$
in other ferromagnetic metals.\cite{Berger} In perovskite manganites, $R_{S}$
is proportional to $\rho _{xx}^{2}$ in the metallic region supporting the
known side-jump picture.\cite{ChunRapid} However, the magnetoresistance of
this 2D system, large even below $T_{c}$, precludes a quantitative analysis
based on $R_{S}$ only. Instead, in Fig. 5, we present the magnetic field
dependence of the Hall conductivity $\sigma _{xy}$ at various temperatures.
Below $T_{min}$, the temperature dependence of the Hall conductivity falls
onto a single curve within the margin of experimental error. For
contributions to the anomalous Hall effect at roughly
temperature-independent magnetization, such a behavior can indeed be
attributed to a side-jump effect. The side-jump contribution in the metallic
conductivity regime, depends on the product of the displacement of the
center of gravity of the wavepacket describing charge carriers, which
depends only on quantum mechanical phases, and the probability of the side
jump. The latter in turn is proportional to the nonequilibrium population of
carriers in the presence of electric field, governed by the momentum
relaxation (transport) time. If the same mechanism, for example scattering
from fluctuations of core spins, leads both to side jumps and to the
transport relaxation time, then the anomalous Hall conductivity is
temperature-independent. We note that the side-jump mechanism is close in
its physical picture to the hopping mechanism of the anomalous Hall effect
that we discussed recently.\cite{ChunPRL,Yuli} The anomalous Hall effect in
the hopping regime arises from topological phases in disordered magnetic
background in the presence of Dzyaloshinskii-Moriya spin-orbit interactions.
The macroscopic mechanism that we proposed explains the scaling behavior of
the Hall effects in La$_{0.67}$(Ca,Pb)$_{0.33}$MnO$_{3}$ single crystals
near $T_{c}$.\cite{ChunPRL} The anomalous Hall contribution due to the
change in the magnetic texture in the regime of the spin glass transition in
localized phase in a certain sense is similar to side-jump effects arising
in the course of scattering by magnetic fluctuations in metallic regime.
Both these mechanisms that are difficult to distinguish in the regime close
to metal-insulator transition may cause the additional $R_{S}$ below $%
T_{min} $ that we observe.

\section{Correspondence between transport studies and neutron scattering data%
}

We now briefly discuss correspondence between our results and neutron
scattering data on magnetic ordering in layered manganites. Neutron
scattering data suggest that the antiferromagnetic superexchange is more
pronounced in the quasi-2D layered system because of reduced dimensionality
and the mixed FM and AFM features are interpreted as canted bilayer states. 
\cite{Kubota,Hirota,Perring,Osborn} However, these results can be
interpreted otherwise, as mentioned by Osborn et al.\cite{Osborn} The first
scenario, i.e. much weaker intrabilayer correlations compared to intraplanar
correlation, is partially true. Chatterji et al. found that the intraplanar
exchange interaction is three times larger than the intrabilayer coupling
although nearest-neighbor distances are almost equal.\cite{Chatterji} The
second possible scenario is an inhomogeneous distribution of FM and AFM
regions. Monte Carlo calculations by Moreo et al.\cite{Moreo} reproduced a
pseudogap feature observed in~\cite{Dessau} in the same system that we
studied. Moreo et al. concluded that this was caused by the formation of FM
metallic clusters in an insulating AFM host. It is noteworthy that the
many-body ground state of the double exchange model has local FM order
without long range order \cite{Zang}. Thus, the interpretation of canted
state based on neutron scattering data is inconclusive. Our observation of
metastability below 40 K favors the cluster model which allows intrinsic
disorder, although the system may transform to the canted state after a long
time relaxation or by external field.\cite{fromGolosov} In any case, La$%
_{1.2}$Sr$_{1.8}$Mn$_{2}$O$_{7}$ is at a delicate balance between competing
magnetic forces. Indeed, when La is partially substituted by Nd, the spin
glass transition is more pronounced.\cite{Moritomo2}

\section{Conclusion}

In conclusion, we studied AC/DC magnetization and longitudinal/transverse
transport properties of La$_{1.2}$Sr$_{1.8}$Mn$_{2}$O$_{7}$ single crystals
in detail. In the low field limit, the magnetic state changes from a
ferromagnet to a spin glass at 40 K where the resistivity starts to
increase. We interpret the resistivity upturn as an indication of a cross
over from a ferromagnetic state to a (disordered) canted state as predicted
by de Gennes. It is supported by the increase of the upturn temperature with
field. This transition is accompanied by additional anomalous Hall effects
due to the change in the magnetic texture while the carrier concentration
deduced from ordinary Hall effects remains constant.

\section{Acknowledgment}

The authors would like to thank M. Jaime and H. Yanagihara for helping
experiments and J. P. Renard for useful comments. This work was supported in
part by DOE DEFG-91ER45439.

%TCIMACRO{
%\TeXButton{Figure 1}{\begin{figure}
%\noindent
%\caption{In-plane resistivity $\rho_{xx}$ (H//$c$) and low field dc magnetizations (H//$ab$) 
%of a ${\rm La_{1.2}Sr_{1.8}Mn_2O_7}$ single crystal 
%as a function of temperature.}
%\label{fig1}
%\end{figure}%
%}}%
%BeginExpansion
\begin{figure}
\noindent
\caption{In-plane resistivity $\rho_{xx}$ (H//$c$) and low field dc magnetizations (H//$ab$) 
of a ${\rm La_{1.2}Sr_{1.8}Mn_2O_7}$ single crystal 
as a function of temperature.}
\label{fig1}
\end{figure}%
%
%EndExpansion
%TCIMACRO{
%\TeXButton{Figure 2}{\noindent
%\begin{figure}
%\caption{Frequency-dependent ac susceptibility (a) and remanent magnetization 
%(b) of the same crystal as a function of temperature. For comparison, ac susceptibility of 
%a ${\rm La_{0.7}Ca_{0.3}MnO_3}$ single crystal is shown together (dashed-line in (a)).
%The inset of (b) shows the logarithmic time dependence 
%of the thermoremanent magnetization decay.}
%\label{fig2}
%\end{figure}%
%}}%
%BeginExpansion
\noindent
\begin{figure}
\caption{Frequency-dependent ac susceptibility (a) and remanent magnetization 
(b) of the same crystal as a function of temperature. For comparison, ac susceptibility of 
a ${\rm La_{0.7}Ca_{0.3}MnO_3}$ single crystal is shown together (dashed-line in (a)).
The inset of (b) shows the logarithmic time dependence 
of the thermoremanent magnetization decay.}
\label{fig2}
\end{figure}%
%
%EndExpansion
%TCIMACRO{
%\TeXButton{Figure 3}{\begin{figure}
%\caption{(a) Hall resistivity $\rho_{xy}$ as a function of field at selected temperatures. 
%The lines are guides for the eyes. (b) Temperature dependences of 
%$n_{eff}$ (open circles) and $R_s$ (solid circles).}
%\label{fig3}
%\end{figure}%
%}}%
%BeginExpansion
\begin{figure}
\caption{(a) Hall resistivity $\rho_{xy}$ as a function of field at selected temperatures. 
The lines are guides for the eyes. (b) Temperature dependences of 
$n_{eff}$ (open circles) and $R_s$ (solid circles).}
\label{fig3}
\end{figure}%
%
%EndExpansion
%TCIMACRO{
%\TeXButton{Figure 4}{\begin{figure}
%\caption{In-plane conductivity $\sigma_{xx}$ shows a square-root $T$ 
%dependence below $T_{min}$ with a field-independent slope. $H$ was applied parallel
%to the $c$-axis.}
%\label{fig4}
%\end{figure}%
%}}%
%BeginExpansion
\begin{figure}
\caption{In-plane conductivity $\sigma_{xx}$ shows a square-root $T$ 
dependence below $T_{min}$ with a field-independent slope. $H$ was applied parallel
to the $c$-axis.}
\label{fig4}
\end{figure}%
%
%EndExpansion
%TCIMACRO{
%\TeXButton{Figure 5}{\begin{figure}
%\caption{Hall conductivity $\sigma_{xy}$ as a function of field below $T_c$.  
%Below $T_{min}$, all data below 4 T collapse onto a single curve. The lines are guides 
%for the eyes.}
%\label{fig5}
%\end{figure}%
%}}%
%BeginExpansion
\begin{figure}
\caption{Hall conductivity $\sigma_{xy}$ as a function of field below $T_c$.  
Below $T_{min}$, all data below 4 T collapse onto a single curve. The lines are guides 
for the eyes.}
\label{fig5}
\end{figure}%
%
%EndExpansion


\begin{references}
\bibitem{intergrowth}  We would like to mention that previous studies show
that there can be some intergrowth of three-dimensional La$_{1-x}$Sr$_{x}$MnO%
$_{3}$ layers in this system \cite{Potter}. Our crystal also has a secondary
ferromagnetic transition at 250 K presumably due to the intergrowth.
However, the relative magnetization was less than 0.6 \%, a too small value
to account for more than the 50 \% reduction in ZFC magnetization at low
temperatures.
\bibitem{transitions}  We assume that doping corresponds to the part of the
phase diagram, in which the ordered state is a ferromagnet. At a different
doping, the high temperature transition can be the
antiferromagnet-to-paramagnet transition, and the low temperature transition
will then be the antiferromagnet-to-canted-phase transition.
\bibitem{compare3D}  This result contrasts with what was observed in La$%
_{1-x}$A$_{x}$MnO$_{3}$, where $n_{eff}$ is several times larger than the
nominal doping concentration and attributed to charge compensation effects. 
\cite{ChunBrief}
\bibitem{fromGolosov}  Even though single-site mean field calculation with
relevant parameters points to the canted phase, the authors of Ref. 8 noted
that the real system might possess more complicated spin ordering like spin
glasses.
\end{references}
\end{document}